# All-Electrical Layer-Spintronics in Altermagnetic Bilayer

Rui Peng,[1,*] Jin Yang,[1,2] Lin Hu[3,4,5] Wee-Liat Ong,[2] Pin Ho,[6] Chit Siong Lau,[1,6]

Junwei Liu[7], Yee Sin Ang[1,*]

[1]Science, Mathematics and Technology (SMT) Cluster, Singapore University of Technology and Design, Singapore 487372

[2]ZJU-UIUC Institute, College of Energy Engineering, Zhejiang University, Jiaxing, Haining, Zhejiang, 314400, China

[3]Centre for Quantum Physics, Key Laboratory of Advanced Optoelectronic Quantum Architecture and Measurement (MOE), School of Physics, Beijing Institute of Technology, Beijing 100081, China

[4]Beijing Key Lab of Nanophotonics & Ultrafine Optoelectronic Systems, School of Physics, Beijing Institute of Technology, Beijing 100081, China

[5]Beijing Computational Science Research Center, Beijing 100193, China

[6]Institute of Materials Research and Engineering (IMRE), Agency for Science, Technology and Research (A*STAR), Singapore 138634

[7]Department of Physics, Hong Kong University of Science and Technology, Hong Kong 999077, China.

*Corresponding authors: rui_peng@sutd.edu.sg; yeesin_ang@sutd.edu.sg

## Abstract

Electrical manipulation of spin-polarized current is highly desirable yet tremendously challenging in developing ultracompact spintronic device technology. Here we propose a scheme to realize the all-electrical manipulation of spin-polarized current in an altermagnetic bilayer. Such a bilayer system can host layer-spin locking, in which one layer hosts a spin-polarized current while the other layer hosts a current with opposite spin polarization. An out-of-plane electric field breaks the layer degeneracy, leading to a gate-tunable spin-polarized current whose polarization can be fully reversed upon flipping the polarity of the electric field. Using first-principles calculations, we show that CrS bilayer with C-type antiferromagnetic exchange interaction exhibits a hidden layer-spin locking mechanism that enables the spin polarization of the transport current to be electrically manipulated via the layer degree of freedom. We demonstrate that sign-reversible spin polarization as high as 87% can be achieved at room temperature. This work presents the pioneering concept of *layer-spintronics* which synergizes altermagnetism and bilayer stacking to achieve efficient electrical control of spin.

# 1. Introduction

Spintronics, a solid-state device concept in which the spin degree of freedom (DOF) of electrons is harnessed for information processing and storage, has received tremendous research interest in the past decades [1-6]. Devices capable of generating pure-spin and spin-polarized currents form the backbones of spintronics technology [1-6]. For ultracompact device integration, the *electrical* generation and detection of the spin current via electric field are highly desirable [7-10]. However, as spin is associated with the angular momentum and magnetic moment, the manipulation of spin signals often relies on magnetic means which are typically not compatible with compact device integration as the stray magnetic fields can influence the operations of neighboring spintronic units [1-6]. The search for an effective electric-field-based control mechanism of spin polarizations remains an open challenge.

Two-dimensional (2D) materials offer a viable platform for developing next-generation solid-state device technology, broadly covering not only conventional computing electronics [11], but also novel device architectures such as spintronics [12], valleytronics [13], and neuromorphic devices [14]. Beyond their monolayer forms, 2D materials can be vertically stacked into bilayers coupled via weak van der Waals forces. Such bilayer stacking generates an additional *layer* DOF as electrons can reside in either the top or the bottom layers [15-30] and are well-separated by a van der Waals gap with a sizable potential barrier. Layer-degeneracy can be electrostatically broken by an out-of-plane electric field generated using an electrical gate. The resulting device concept, i.e. *layertronics*, harnesses the layer DOF for information processing [31], and is inherently more compatible with direct electrical manipulation. Remarkably, through bilayer stacking, DOF previously not easily accessible using electric fields, such as spin DOF, can now be coupled with the layer DOF to introduce electrical control of spin DOF [27-30].

In this work, we propose a scheme to generate, tune, and reverse spin-polarized currents by combining the concepts of *layertronic bilayers* with *altermagnetism*. In altermagnets – an unconventional antiferromagnetic system with strong potential in spintronics, thermoelectrics, transistors, multiferroics, and superconducting device applications [32-45], the spin-momentum coupled band structure is usually accompanied by noncollinear spin currents [32,33]. By constructing an altermagnetic bilayer with *layer-contrasting* altermagnetic bands in which the two layers host *opposite* spin-split band structures in the momentum space, a peculiar altermagnetic-enabled layer-spin locking effect can arise. An out-of-plane electric field breaks the layer degeneracy and the accompanying spin degeneracy, thus leading to a gate-tunable spin polarization whose polarity can be flipped upon reversing the polarity of the gate voltage.

Using first-principles density functional theory (DFT) simulations, we demonstrate that CrS is a candidate 2D material to realize our proposed scheme. CrS monolayer is an altermagnet with crystal-symmetry-paired spin-valley locking (CSVL) in the absence of spin-orbit coupling. We show that CrS bilayer with C-type antiferromagnetic (AFM) exchange interaction exhibits a *hidden* layer-contrasting CSVL. Such layer-contrasting CSVL offers a layer-spin locking mechanism that enables the spin polarization to be electrically manipulated by an out-of-plane electric field. Sign-reversible spin polarization as large as 87% can be generated at room temperature under practical electrostatic gating in CrS bilayer. Our findings concretely establish a pioneering route in utilizing altermagnetic bilayer as a viable building block of all-electrical spintronic devices and may serve as a harbinger of *layer-spintronics* – a novel device concept where layer and spin DOFs are intertwined.

## 2. Methods

First-principles calculations are performed based on DFT [46] as implemented in the Vienna ab initio simulation package (VASP) [47]. Exchange-correlation interaction is described by the Perdew-Burke-Ernzerhof (PBE) parametrization of generalized gradient approximation (GGA) [48]. DFT+U method is employed to describe the strong correlations of the Cr-3*d* orbitals [49]. We considered effective Hubbard $U_{eff}$ = 0, 1, 2, 3 and 4 eV, and adopted $U_{eff}$ = 3 because the calculated results are consistent with those of hybrid functional HSE06 [50] (**Fig. S1**). All structures are relaxed until the force on each atom is less than 0.01 eV/Å. The cutoff energy and electronic iteration convergence criterion are set to 400 eV and $10^{-5}$ eV, respectively. To sample the 2D Brillouin zone, a Monkhorst–Pack (MP) k-grid mess of 9 × 9 × 1 is used [51]. To avoid the interaction between adjacent layers, a vacuum slab of 20 Å is added. DFT-D3 method is employed to treat the van der Waals interaction [52]. Phonon calculations are carried out using the PHONOPY code [53]. Ab initio molecular dynamics (AIMD) simulations are performed at 300 K for 5 ps with a time step of 1 fs [54]. The spin-resolved transport properties are calculated using a housing-made code, in which the electron energy and electron group velocity are evaluated from the Wannier-based tight-binding Hamiltonian [55].

## 3. Results and discussion

### 3.1 Layer-contrasting spin-polarized current

CSVL, where electronic states in different valleys have opposite spin splitting due to the crystal symmetry ensuring the exactly zero magnetization, is one the intrinsic feature of altermagnet and leads to the unique unconventional piezomagnetism [33] and noncollinear spin current generation [32,33] in altermagnet [**Fig. 1(a)**]. By stacking two altermagnetic monolayers into a bilayer, layer-contrasting CSVL can be realized. In general, the following symmetry requirements must be satisfied:

(i) The intralayer opposite-spin sublattices are connected by rotation ***R*** (proper or improper, symmorphic or nonsymmorphic) and not by translation ***t*** or inversion ***P***. ***RT*** gives rise to $E_{\uparrow top/bot}(k) = E_{\downarrow top/bot}(Rk)$, and thus the spin splitting at k and ***R***k should exhibit the opposite sign; and

(ii) The interlayer opposite-spin sublattices are connected by translation ***t*** or inversion ***P***. ***PT*** or ***t*$_{1/2}$*T** (***t*$_{1/2}$** is a half unit cell translation) guarantees the global spin splitting is zero, i.e., $E_{\uparrow(top+bot)}(k) = E_{\downarrow(top+bot)}(k)$. Thus, the spin splitting from two layers should cancel out with each other, forming the hidden layer-contrasting CSVL physics, i.e., $E_{\uparrow top}(k) = E_{\downarrow bot}(k)$, $E_{\downarrow top}(Rk) = E_{\uparrow bot}(Rk)$.

An altermagnetic monolayer can generate a spin-polarized current [32,33] [**Fig. 1(b)**]. In this work, we show that in an appropriately stacked bilayer, *layer-contrasting* altermagnetic spin-polarized current can be induced such that the top and bottom layers host charge current of opposite spin polarizations [**Fig.1(c)**]. In the absence of electric field, due to layer degeneracy, the spin polarization of the two layers mutually cancels out each other, leading to zero net polarization. An external out-of-plane electric field breaks the layer degeneracy [**Fig. 1(d)**] and causes the electronic bands of the top and bottom layers to energetically shift away from each other. Such layer degeneracy breaking causes the spin-polarized current of one layer with energy bands closer to the Fermi level to dominate the overall carrier conduction, thus generating finite spin polarization. Intriguingly, when the electric field is reversed, the other layer with opposite spin polarization dominates the charge transport, causing the spin-polarization of the transport current to undergo a sign reversal [**Fig. 1(e)**]. Electric-field-driven sign-reversible spin-polarized current can thus be realized in such an altermagnet bilayers, which is uniquely enabled by a previously unexplored *layer-spin* locking mechanism in a bilayer heterostructure.

### *3.2 CSVL in CrS monolayer*

We now demonstrate that CrS [38] provides a potential material platform to realize the mechanism of gate reversible spin polarization introduced above (**Fig. 2**). CrS monolayer has a space group of P4/mmm [**Fig. 2(a)**]. The lattice constant and layer thickness of CrS monolayer are *a* = *b* = 3.60 Å and *c* = 3.13 Å, respectively. The dynamic and thermal stabilities of CrS monolayer are confirmed by vibrational phonon calculations and AIMD simulations (**Fig. S2**). The calculated magnetic moment on each Cr atom is 3.44 $\mu_B$. We consider three different magnetic configurations for CrS monolayer: AFM-Néel, AFM-Stripy and FM (**Fig. S3**). It is found that the energy of AFM-Néel is 0.48 and 1.30 eV/unit cell lower than AFM-Stripy and FM, respectively, indicating that AFM-Néel is the magnetic ground state. In this case, the two opposite-spin Cr atoms are related by a mirror symmetry $M_{110}$.

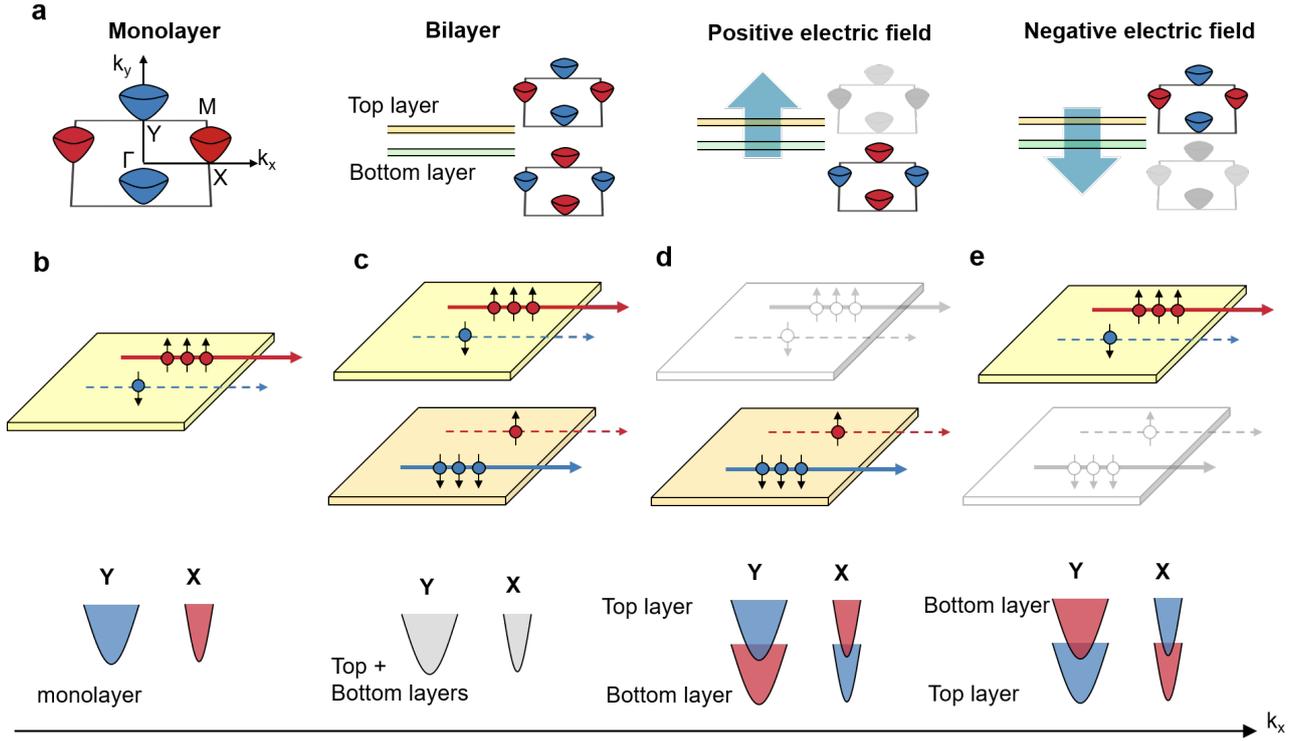

**Fig. 1. General scheme of field-tunable and reversible spin polarization in altermagnetic bilayer.** (a) Schematics of CSVL in altermagnet monolayer and layer-locked CSVL in altermagnet bilayer. (b) Schematics of spin current and band structure along $k_x$ direction in altermagnet monolayer. (c) Schematics of layer-locked spin current and band structure along $k_x$ direction in altermagnet bilayer. (d) Schematics of layer-polarized spin current and band structure along $k_x$ direction in altermagnet bilayer under positive electric field. (e) The same as (d) but under negative electric field.

The band structure of CrS monolayer shows a direct band gap of 0.46 eV located at the X and Y valleys [**Fig. 2(b)**], forming two pair of valleys at the lowest conduction and highest valence bands. Due to the absence of ***PT*** and $t_{1/2}T$ symmetry, the unconventional spin splitting can be observed in the reciprocal momentum space. The spin splitting is opposite at mirror symmetry-related points, which reflects the same mirror symmetries as the two opposite-spin sublattices. The alternating spin polarizations in both direct physical space and reciprocal momentum space suggest the presence of altermagnetism in CrS monolayer [**Fig. 3(a)**]. Importantly, the spin and valley DOFs are locked together at the X and Y valleys, thus forming the CSVL where electronic states at one valley is primarily composed of only one spin [**Fig. 2(b)**]. As the X/Y valleys are energetically degenerate, the spin-up and spin-down density of states are equal, which is consistent with AFM coupling.

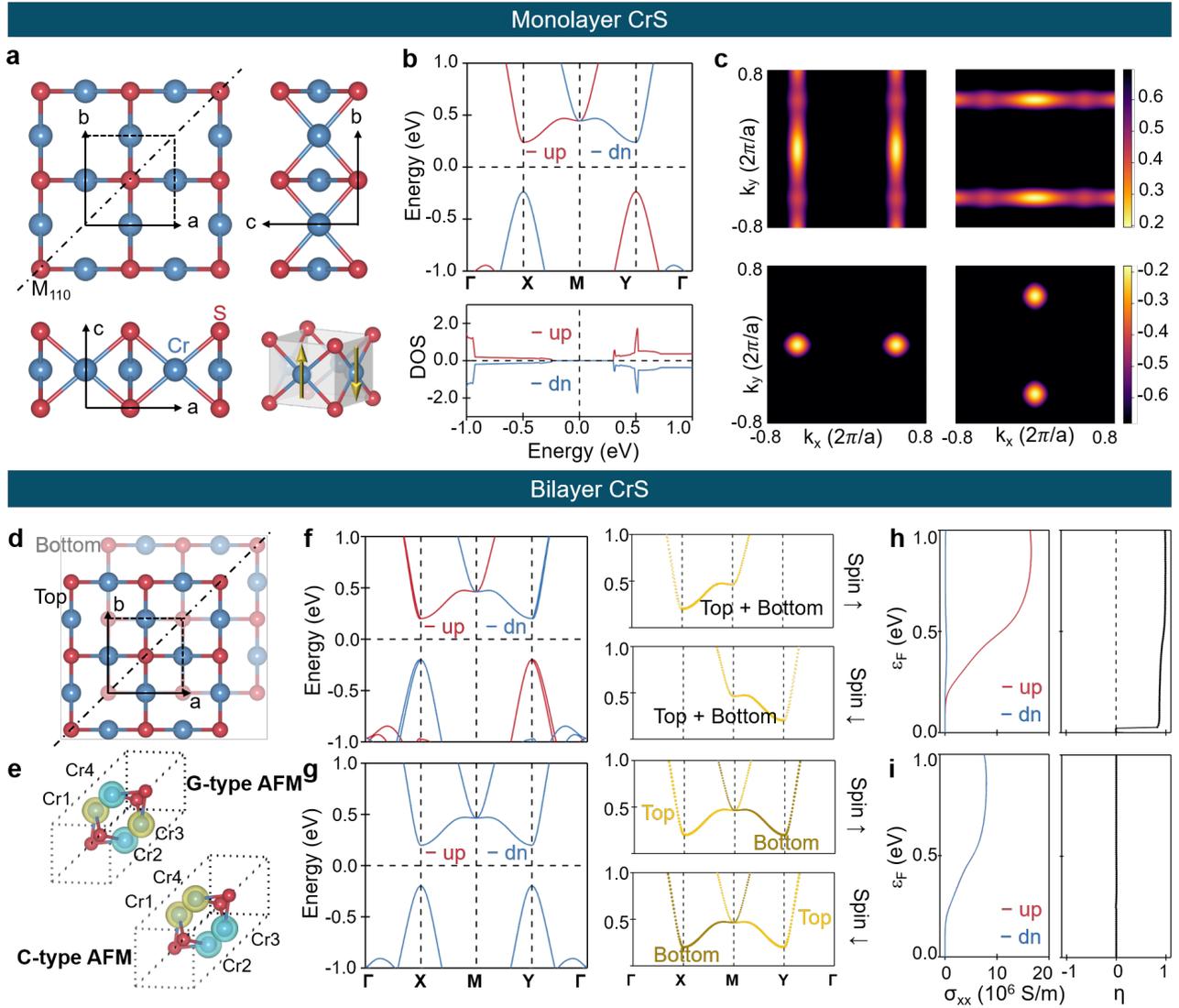

**Fig. 2. Electronic structures, spin-valley coupling, and layer degree of freedom in CrS.** (a) Crystal structures of CrS monolayer and coordination environment of Cr atom. The yellow arrows indicate the spin direction of Cr atoms. (b) Spin-polarized band structure (top) and density of states (bottom) of CrS monolayer. The Fermi level is set to the middle of the band gap. (c) Energy contour of the lowest conduction band [CBM, CBM + 0.5 eV] (top) and highest valence band [VBM - 0.5 eV, VBM] (bottom) of CrS monolayer around X (left) and Y (right) valleys. (d) Crystal structure of CrS bilayer. (e) Spin charge density of G-type AFM (top) and C-type AFM (bottom). The yellow and blue isosurfaces indicate spin-up and spin-down density, respectively. (f) Spin-polarized band structure and orbital-resolved band structure of G-type AFM. The Fermi level is set to the middle of the band gap. (g) Spin-polarized band structure and orbital-resolved band structure of C-type AFM. The Fermi level is set to the middle of the band gap. (h) Spin-resolved conductivity and spin polarization of the transport current in G-type AFM. (i) Spin-resolved conductivity and spin polarization of the transport current in C-type AFM.

The energy contours of the lowest conduction and the highest valence bands in reciprocal space [**Fig. 2(c)**] reveal exceptionally strong anisotropy in the X/Y valleys of the conduction bands and isotropy in the X/Y valleys of the valence bands. In the lowest conduction band, the energy band around X valley disperses steeply along the x-direction while those around Y valley disperses relatively gradually along the x-direction. We thus expect the conduction bands to generate strong altermagnetic noncolinear spin-polarized current. Using the Boltzmann equation under the constant relaxation time approximation, the 2D electrical conductivity mediated by electrons residing in the X or Y valleys is [56]:

$$\sigma_{ij}^V = -\frac{\tau e^2}{4\pi^2}\int d^2k \frac{\partial f(E_{\boldsymbol{k}}^V)}{\partial E_{\boldsymbol{k}}^V} v_{\boldsymbol{k}_i}^V v_{\boldsymbol{k}_j}^V$$

(1)

where the superscript V = X, Y denotes the X and Y valleys, respectively, i and j denotes two orthogonal directions, $\tau$ is the carrier relaxation time, $E_{\boldsymbol{k}}^V$ is the energy dispersion around the V valley, $v_{\boldsymbol{k}_i}^V = \frac{\partial E_{\boldsymbol{k}}^V}{\partial k_i}$ and $v_{\boldsymbol{k}_j}^V = \frac{\partial E_{\boldsymbol{k}}^V}{\partial k_j}$ is the velocity along *i* and *j* directions, and $f(E_{\boldsymbol{k}}^V)$ is the Fermi-Dirac distribution function.

By fitting the energy band around the conduction band minima (CBM) with a parabolic energy dispersion, the mirror symmetry $M_{110}$ of the system allows the energy dispersion of the two valleys to be written as:

$$E_{\boldsymbol{k}}^X = \frac{\hbar^2}{2m_1}k_x^2 + \frac{\hbar^2}{2m_2}k_y^2$$

(3)

$$E_{\boldsymbol{k}}^Y = \frac{\hbar^2}{2m_1}k_y^2 + \frac{\hbar^2}{2m_2}k_x^2$$

(4)

where $m_1$ and $m_2$ are the effective masses of electrons around X (Y) point along x and y (y and x) directions, respectively. The x-directional electrical conductivities at the X and Y valleys becomes (see more details in **Supplementary Materials**):

$$\sigma_{xx}^X = \frac{ne^2\tau}{m_1}$$

(4)

$$\sigma_{xx}^Y = \frac{ne^2\tau}{m_2}$$

where $n = \frac{\sqrt{m_1 m_2}}{2\pi\hbar^2} \int_0^\infty E_{\boldsymbol{k}}^X (-\frac{\partial f}{\partial E_{\boldsymbol{k}}^X}) dE_{\boldsymbol{k}}^X$. Here $\sigma_{xx}^X \neq \sigma_{xx}^Y$ as a result of the electron effective mass difference of the altermagnetic energy bands. Furthermore, as the valley and spin are locked together [**Figs. 2(b)** and **3(a)**], the conductivities at each valley is spin polarized, i.e. $\sigma_{xx}^X = \sigma_{xx}^{X(\uparrow)}$ and $\sigma_{xx}^Y = \sigma_{xx}^{Y(\downarrow)}$ where the up/down arrows in the superscript denotes spin-up/down states, respectively. The total charge current ($\sigma_t$) along the x-direction is thus:

$$\sigma_t = \sigma_{xx}^{X(\uparrow)} + \sigma_{xx}^{Y(\downarrow)}$$

(6)

with a finite spin polarization:

$$\eta = \frac{\sigma_{xx}^{X(\uparrow)} - \sigma_{xx}^{Y(\downarrow)}}{\sigma_{xx}^{X(\uparrow)} + \sigma_{xx}^{Y(\downarrow)}}$$

(7)

The conductivity and the spin polarization of CrS monolayer along x direction are shown in **Fig. S4**. Over the Fermi level ($E_F$) range of [0, 1] eV, the spin polarization, arising from the altermagnetic nature of CrS, can reach as large as 97%.

### *3.3 Layer-contrasting* spin-polarized current *in CrS bilayer*

To generate layer DOF, two CrS monolayers are vertically stacked into a CrS bilayer. The bottom layer has a 1/2(*a+b*) translation relative to the top layer [**Fig. 2(d)**]. The dynamic stability of CrS bilayer is confirmed by vibrational phonon calculations (**Fig. S5**). Two magnetic configurations, i.e. G-type AFM and C-type AFM, are considered [**Fig. 2(e)**] [57]. The total energy of the G-type AFM configuration is only 0.002 eV/unit cell more stable than that of the C-type AFM, thus suggesting the two configurations are almost energetically equivalent.

For G-type AFM configuration, the intralayer opposite-spin sublattices [i.e. the pairs of Cr1 and Cr2, and of Cr3 and Cr4 as labelled in **Fig. 2(e)**] are connected by mirror symmetry $M_{110}$, and the interlayer opposite-spin sublattices [i.e. the pairs of Cr1 and Cr3, and of Cr2 and Cr4 as labelled in **Fig. 2(e)**] are connected by roto inversion symmetry $\bar{4}$. CSVL is preserved in the G-type AFM configuration. As shown in **Fig. 2(f)**, the band structure of G-type AFM configuration resembles that of CrS monolayer with a bandgap of 0.40 eV. The carrier conduction can be generalized to the bilayer cases. In this case, the *layer-polarized*

valley-dependent electrical conductivity becomes

$$\sigma_{ij,L}^{V} = -\frac{\tau e^2}{4\pi^2} \int d^2k \frac{\partial f(E_{k,L}^{V})}{\partial E_{k,L}^{V}} v_{k_i}^{V} v_{k_j}^{V}$$

(8)

where $L = \pm$ denotes the top and bottom layers, respectively. Consider the x-directional carrier conduction, because of the CSVL, the CBM of X valley is dominated by the spin-up states [**Fig. 2(f)**], while that of Y valley is dominated by the spin-down states, yielding $\sigma_{xx,L}^{X} = \sigma_{xx,L}^{X(\uparrow)}$ and $\sigma_{xx,L}^{Y} = \sigma_{xx,L}^{Y(\downarrow)}$ for G-type AFM configurations. Due to the layer degeneracy, we have $\sigma_{xx,+}^{X(\uparrow)} = \sigma_{xx,-}^{X(\uparrow)}$, and $\sigma_{xx,+}^{Y(\downarrow)} = \sigma_{xx,-}^{Y(\downarrow)}$. The total current is:

$$\sigma_t = \underbrace{\sigma_{xx,+}^{X(\uparrow)} + \sigma_{xx,+}^{Y(\downarrow)}}_{top\ layer} + \underbrace{\sigma_{xx,-}^{X(\uparrow)} + \sigma_{xx,-}^{Y(\downarrow)}}_{bottom\ layer}$$

(9)

which carriers a spin polarization:

$$\eta = \frac{\sigma_{xx,+}^{X(\uparrow)} + \sigma_{xx,-}^{X(\uparrow)} - \sigma_{xx,+}^{Y(\downarrow)} - \sigma_{xx,-}^{Y(\downarrow)}}{\sigma_{xx,+}^{X(\uparrow)} + \sigma_{xx,+}^{Y(\downarrow)} + \sigma_{xx,-}^{X(\uparrow)} + \sigma_{xx,-}^{Y(\downarrow)}}$$

(10)

Here $\sigma_{xx,L}^{V,(S)}$ is computed fully numerically from the DFT calculated band structures. Carrier conduction is thus similar to the case of CrS monolayer, except that the transport current is now *duplicated* by the two layers with the same spin polarization in each layer [**Figs. 2(h) and 3(b)**].

The C-type AFM configuration differs dramatically from the G-type AFM counterpart. Under the C-type AFM configuration, the intralayer opposite-spin sublattices (Cr1 and Cr2, Cr3 and Cr4) are connected by mirror symmetry $M_{110}$, while the interlayer opposite-spin sublattices (Cr1 and Cr4, Cr2 and Cr3) are connected by inversion symmetry **P**. Due to the presence of **PT** symmetry, the global band structure of C-type AFM shows no spin splitting [**Fig. 2(g)**].

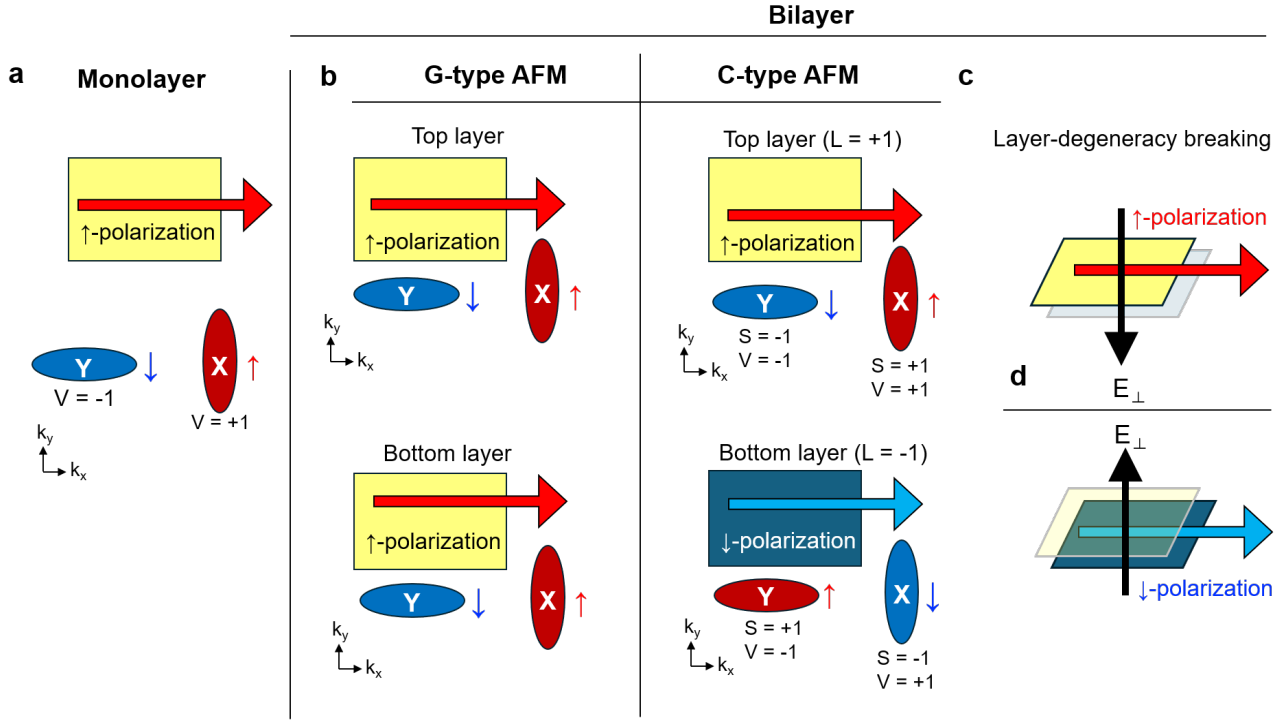

**Fig. 3. Schematic illustrations of layer-locked spin polarization in CrS bilayer.** (a) Altermagnetic spin-polarized current in CrS monolayer. (b) Altermagnetic CrS bilayer under the G-type AFM configuration. Both top and bottom layers host the same altermagnetic spin-split band structures and carry a conduction current with the same spin polarization. (c) Same as (b) but under C-type AFM configuration. The top and bottom layer host altermagnetic band structures of opposite spin-splitting. The spin, layer and valley indices are S = ±1, L = ±1, and V = ±1, respectively. To preserve the layer-spin-valley coupling condition of L·S·V = 1, the spin polarization of the top and bottom layer is opposite to each other. The total transport current carry zero spin polarization. (d) An out-of-plane electric field breaks the layer degeneracy in C-type AFM CrS bilayer, leading to a transport current with finite spin polarization. Spin polarization can be flipped by changing the polarity of the external electric field.

Unlike the case of G-type AFM configuration where the two layers are identical copies, the CSVL in the C-type AFM case is *layer-contrasting*. Such layer-contrasting CSVL is captured by the condition $S \cdot V \cdot L = 1$, where S = ±1 represent spin-up and spin-down states, V = ±1 represent X and Y valleys, and L = ±1 represent top and bottom layers, respectively. Importantly, the CSVL condition dictates that two of the three indices must change signs simultaneously. Consider the top layer of L = +1 [see **Fig. 3(c)** for schematic illustrations], to fulfil $S \cdot V \cdot L = 1$, the spin-up (S = +1) electron must resides in the X valley (V = +1) while the spin-up (S = -1) electron must reside in the Y valley (V = -1). In the bottom layer (L = -1), the spin-valley combination must be interchanged to preserve $S \cdot V \cdot L = 1$, i.e. spin-up (S = +1) electron resides in the Y valley (V = -1) while spin-down electron (S = -1) resides in the X valley (V = +1). Such a layer-contrasting CSVL leads to a layer-spin locking effect where the top layer carries an electrical current of one spin polarization while the bottom layer carries an oppositely spin-polarized current. Accordingly, the spin polarization becomes:

$$\eta = \frac{\sigma_{xx,+}^{X(\uparrow)} + \sigma_{xx,-}^{Y(\uparrow)} - \sigma_{xx,+}^{Y(\downarrow)} - \sigma_{xx,-}^{X(\downarrow)}}{\sigma_{xx,1}^{X(\uparrow)} + \sigma_{xx,1}^{Y(\downarrow)} + \sigma_{xx,-1}^{X(\downarrow)} + \sigma_{xx,-1}^{Y(\uparrow)}}$$

(11)

The layer-contrasting spin polarization is opposite between the two layers and thus cancels out each other overall (i.e. $\sigma_{xx,+}^{X(\uparrow)} = \sigma_{xx,-}^{X(\downarrow)}$, and $\sigma_{xx,+}^{Y(\uparrow)} = \sigma_{xx,-}^{Y(\downarrow)}$), yielding zero spin polarization [**Fig. 2(i)**]. Nevertheless, the layer-degeneracy of CrS bilayer can be readily broken by an out-of-plane electric field, resulting in electrical manipulation of spin polarization in the transport current [**Fig. 3(d)**].

### *3.4 Electric field reversible layer-spin-polarized current*

The presence of an out-of-plane electric field breaks the layer degeneration, leading to a finite layer polarization at the CBM and valence band maxima (VBM), which can be, respectively, quantified as $\Delta\text{CBM} = E_{top}^{CBM} - E_{bot}^{CBM}$ and $\Delta\text{VBM} = E_{top}^{VBM} - E_{bot}^{VBM}$, where $E_{top}^{CBM/VBM}$ and $E_{bot}^{CBM/VBM}$ are the CBM/VBM band edge energies of the top and bottom layers, respectively (**Fig. 4**). The $\Delta c$ and $\Delta v$ of both G-type AFM and C-type AFM are nearly linearly dependent on the out-of-plane electric field, and exhibit a wide tuning range greater than 300 meV within the electric field strength window of ± 0.15 V/Å [**Figs. 4(a)** and **4(c)** for G-type AFM and C-type AFM configurations, respectively], which is within the dielectric breakdown limit of commonly used gate dielectric SiO$_2$ (i.e. 15 MV/cm) [58].

The spin-polarized and layer-polarized band structures of G-type AFM and C-type AFM under an out-of-plane electric field are shown in **Figs. 4(b)** and **4(d)**, respectively. For G-type AFM configuration, changing the polarity of the electric field switches the layer polarization, i.e. the transport current is dominantly carried by the top layer at negative applied electric field [see 1st column of **Fig. 4(b)**] but can be switched to the bottom layer at positive applied electric field [see 3rd column of **Fig. 4(b)**]. The spin polarization, however, remains unchanged, since both top and bottom layer host identical spin-split bands with the same spin polarization.

Interestingly, in the case of C-type AFM, changing the polarity of the out-of-plane electric field switches *both* the layer and spin polarizations of the transport current. Similar to the case of G-type AFM configuration, the transport current is dominantly localized on the top (bottom) layer under a negative (positive) out-of-plane electric field [see 1st and 3rd column of **Fig. 4(d)** for negative and positive electric field, respectively]. However, because the top and bottom layers carry opposite CSVL, the accompanying spin polarization of the transport current is reversed upon changing the polarity of the out-of-plane electric field, thus leading to an electrically tunable and sign-reversible spin-polarized current in C-type AFM CrS bilayer.

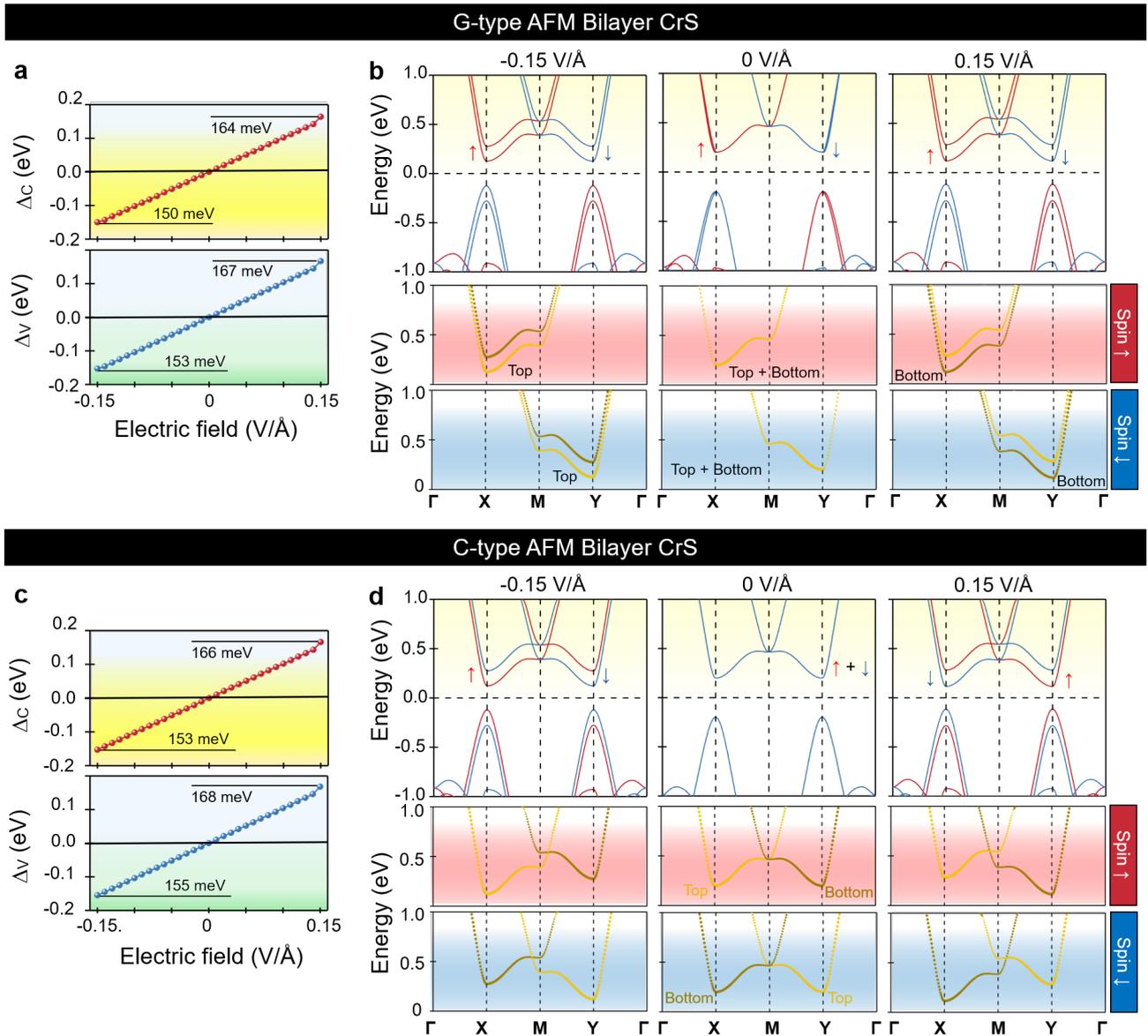

**Fig. 4. Layer-degeneracy breaking of bilayer CrS under an external out-of-plane electric field.** (a) Layer polarizarion at the CBM and VBM as a function of the out-of-plane electric field in G-type AFM. (b) Spin-polarized band structures and orbital-resolved band structures of G-type AFM under the out-of-plane electric field of -0.15, 0, and 0.15 V/Å. The Fermi level is set to the middle of the band gap. (c) Layer polarizarion at the CBM and VBM as a function of the out-of-plane electric field in C-type AFM. (d) Spin-polarized band structures and orbital-resolved band structures of C-type AFM under the out-of-plane electric field of -0.15, 0, and 0.15 V/Å. The Fermi level is set to the middle of the band gap.

In **Fig. 5**, we consider a spin polarizer based on a field-effect transistor device geometry using C-type AFM CrS bilayer as the channel material. The key operation of the device relies on the field-effect tuning of the layer-polarized spin-valley gap $\Delta_L^{X/Y}$ [see **Fig. 5(a)** for the

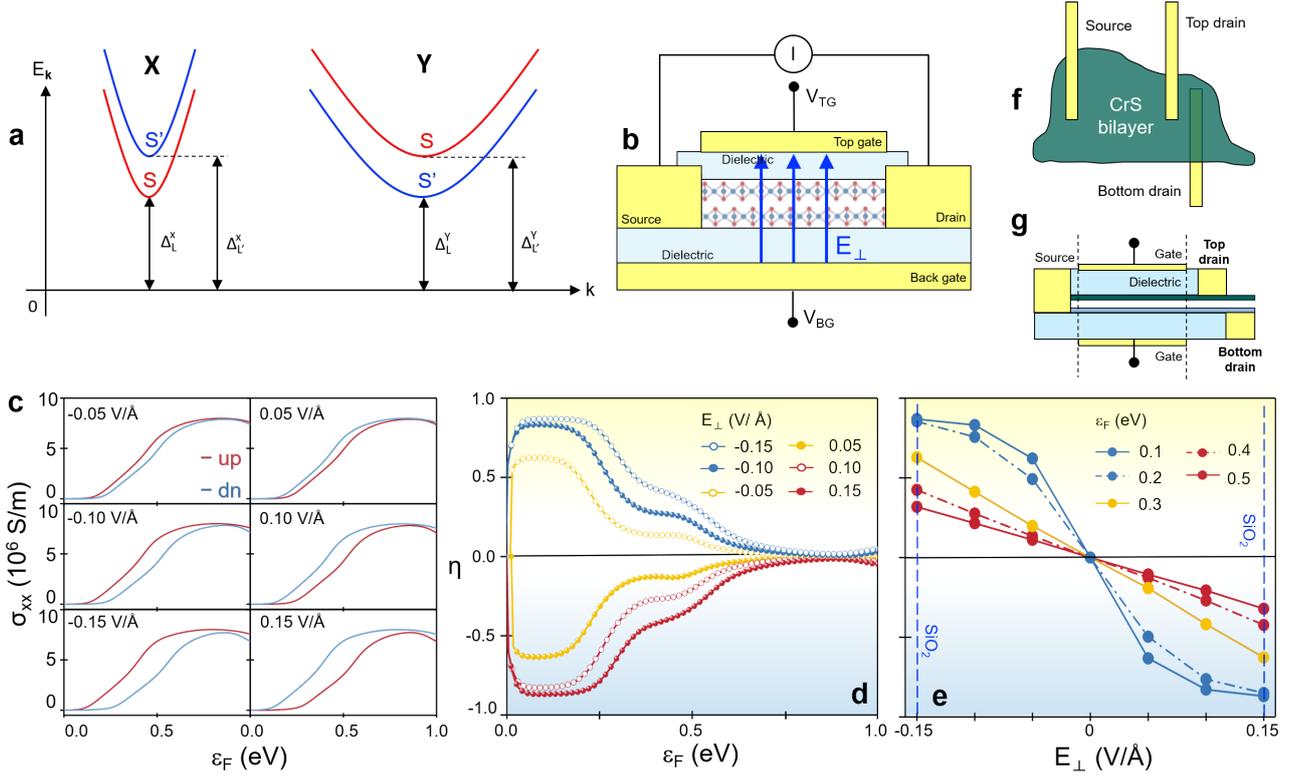

**Fig. 5. Spin polarizer device performance of CrS bilayer.** (a) Band diagram of C-type AFM CrS bilayer showing the layer-spin subbands at the X and Y valleys arising from the layer-degeneracy breaking induced by an out-of-plane electric field. The two layers (indicated by layer index L and L') host contrasting spin-valley coupling (S and S' for spin index; X and Y denotes the two valleys). The layer-dependent band gaps are denoted by $\Delta_L^{X/Y}$ for X/Y valley, respectively. (b) Schematic drawing of a spin-polarized implemented with a field-effect transistor device setup. (c) Spin-resolved conductivity under several out-of-plane electric fields, ±0.05, ±0.10, and ±0.15 V/Å. Spin polarization of the transport current as a function of (d) Fermi level and (e) out-of-plane electric field. The vertical dashed lines indicate the dielectric breakdown limit of $SiO_2$. (f) and (g) show the top and cross-sectional view of a device schematic for layer-polarization read-out. The carrier is injected via an edge-contacted source electrode where CrS bilayer is layer-degenerate. The channel section breaks the layer-spin degeneracy creates layer-spin polarized current. The top and bottom drains allow the layer-polarized current to be separately read out.

band diagram], which is extracted from the DFT band structure calculations (**Fig. 4**). The layer-contrasting band gaps leads to the presence of majority and minority layer-spin subbands which are akin to the majority and minority spin subbands in a ferromagnetic semiconductor. We consider a dual-gated device setup, [see **Fig. 5(b)** for a schematic drawing] which allows both the out-of-plane electric field and the Fermi level of the device to be efficiently controlled [59]. The conductivity is calculated numerically based on the DFT calculated band structures at a temperature of 300 K under several out-of-plane electric field of ±0.05, ±0.10, and ±0.15 V/Å in **Fig. 5(c)**. Such electric field strengths are experimentally achievable through common dielectrics such as $SiO_2$ and $HfO_2$ [60,61]. When the out-of-plane electric field is negative and positive, the spin -polarized conductivities are $\sigma_{xx}^{X(\uparrow)} >$

$\sigma_{xx}^{Y(\downarrow)}$ and $\sigma_{xx}^{Y(\uparrow)} < \sigma_{xx}^{X(\downarrow)}$, respectively, within the Fermi level window of [0, 1] eV, thus revealing the sign-reversal of spin polarization upon switching the polarity of the out-of-plane electric field. Alternative to the fully numerical evaluation of the conductivity and spin polarization using DFT calculated band structure data, we have also developed a semi-analytical transport model based on the parabolic energy dispersion fitting of the CBMs of the layer-spin subbands at X and Y valleys (**Figs. S6** and **S7**). The spin polarization calculated using such semi-analytical model agrees well with the full numerical approach. The spin polarizations as a function of Fermi level and the out-of-plane electric field [**Figs. 5(d)** and **5(e)**] clearly demonstrate the electrical manipulation of spin polarization of the transport current in C-type AFM CrS bilayer. The spin polarization is suppressed when the Fermi level is much higher than band gaps since both layer-spin subbands can jointly contribute to the transport current. The spin polarization reaches up to ±87% at an electric field of ±0.15 V/Å with the Fermi level is set to 0.1 eV above the mid-gap level, thus suggesting the capability of CrS bilayer as a building block of all-electrical layer-spintronics.

It should be noted that the transport current is not only spin-polarized, but also layer-polarized due to the layer-spin coupling. The layer-spin polarization can thus be alternatively read out via the layer localization of the transport current, in addition to magnetically assessing the spin polarization such as using ferromagnetic contacts [62-64]. In **Figs. 5(f)** and **5(g)**, a device setup is proposed for detecting the layer-polarization of the transport current. The carrier is first injected into the layer-degenerate CrS bilayer via the drain electrode under a one-dimensional edge contact geometry [65]. The channel region is electrically gated, thus producing layer-spin polarized current. The drain region is composed of a top electrode contacting the top sublayer and a separate bottom electrode contacting the bottom sublayer, which enables the electrical current spatially localized on each layer to be separately assessed. We note that for the configuration shown in **Fig. 5(g)**, as the layer-polarized current needs to travel an additional distance before reaching the bottom electrode, the interlayer carrier scattering could potentially reduce the layer-polarization detection efficiency. Due to the presence of a van der Waals gap in a bilayer heterostructure, a potential barrier with sizable barrier height and width is present, which can weaken the interlayer coupling. The interlayer scattering effect in bilayer heterostructure is expected to be strongly suppressed as experimentally observed in MoS$_2$ bilayer experimentally [66,67]. Recent experiment has demonstrated the scaling of contact length down to 12 nm in MoS$_2$ monolayer transistor [68]. We thus do not expect the interlayer scattering over a top electrode contact length of around 12 nm, assuming that similar electrode fabrication method of MoS$_2$ can be employed for CrS, to pose severe challenges in observing the layer-polarized current using the two-drain setup in **Fig. 5(g)**.

Finally, we remark that the scheme proposed here is drastically different from the recently proposed spin-polarization control scheme in Ca(CoN)$_2$ and its expanded family of metal-decorated transition metal nitride monolayers [42], which is based on the concept of "*sublayer*" polarization where the electronic states are localized either on the upper or lower sublayers of the same monolayer depending on their spin polarization. Achieving such sublayer-spin coupling would also require a thick monolayer morphology so to ensure that the sublayer polarization is spatially unmixed and resilient against inter-sublayer scattering effect. In contrast, the scheme proposed in this work arises from the *true* layer polarization which is inherently spatially well-separated into two different monolayers constituting the bilayer heterostructure. The presence of a van der Waals gap in the bilayer heterostructure also provide a protection mechanism to reduce interlayer scattering [66,67], thus better preserving the layer-spin polarization as compared to the sublayer polarization case of a monolayer system. The layer-contrasting subbands energy splitting of monolayer [42] and bilayer is approximately related to $eE_\perp d_{ML}$ and $eE_\perp d_{BL}$, respectively, where $d_{ML}$ and $d_{BL}$ are the thicknesses of the monolayer and of the bilayer heterostructure, respectively. As $d_{BL} > d_{ML}$, a larger layer energy splitting can be achieved in a bilayer setup with lower electric field requirement. For example, CrS bilayer can achieve an energy layer-spin splitting over 150 meV with an electric field strength of 0.15 V/Å, which is substantially larger than the 60 meV and 80 meV splitting of Ca(CoN)$_2$ for CBM and VBM [42], respectively, at the same electric field strength. The layer-spin polarization proposed in this work is also less restrictive in terms of symmetry class requirement as compared to the case of sublayer-spin polarization in a monolayer, and thus can be realized in a broader combination of altermagnetic bilayer as long as the stacking preserves the layer-spin locking. Experimentally, CSVL materials with altermagnetism have been demonstrated, so we expect those systems may serve as an alterantive platform to explore the altermangetic bilayer spintronics proposed in this work. We also remark that the scheme relies on volatile control of the spin current. By replacing electric field with ferroelectric substrates, it is also possible to realize electrical control of the spin current in a nonvolatile way [69].

## 4. Conclusion

In summary, a scheme to electrically manipulate and reverse the spin-polarization of electrical transport currents in altermagnet bilayers was proposed in which the spin is interlocked with layer degree of freedom. Using first-principles calculations, we showed that the altermagnetic CrS provides a material platform to realize the proposed layer-spin locking via a layer-contrasting spin-valley coupling effect in their bilayer heterostructures. Gate-tunable current with spin polarization as high as 87% can be achieved and sign-reversed under experimentally achievable gating setup. Due to the layer-spin locking nature of the transport current, the layer-spin polarization can also be assessed using spatially separated top and bottom electrodes in addition to magnetic-based spin polarization read out. Our findings reveal a previously unexplored layer-spin locked mechanism to manipulate spin degree of freedom all electrically in an altermagnetic bilayer setup. These results open up the new concept of layer-spintronics in which electrical control is re-enabled by intertwining spin with layer degree of freedom. The demonstration of altermagnetic bilayer to achieve electrical spin manipulation enriches the device application scenario of altermagnets [70-72].

## Data availability

The data that support the findings of this study are available from the corresponding author upon request.

## Acknowledgements

This work is supported by the Singapore Ministry of Education Academic Research Fund (AcRF) Tier 2 Grant under the award number MOE-T2EP50221-0019 and the SUTD Kickstarter Initiatives (SKI 2021_04_01). J. Y. acknowledges the support of SUTD-ZJU IDEA Visiting Professor Grant (SUTD-ZJU (TR) 202203). C.S.L. acknowledges support from MTC YIRG grant No. M21K3c0124 and MTC IRG grant No. M23M6c0103. P. Ho acknowledges the support from RIE2025 Manufacturing, Trade and Connectivity (MTC) Individual Research Grant (Grant No. M23M6c0101) and Career Development Fund (Grant No. C210812017), administered by A*STAR.

## Competing interests

The authors declare no competing interests.